\documentclass[aps,prd,twocolumn,preprintnumbers,showpacs,nofootinbib,amssymb]{revtex4}
\usepackage{graphicx}
\usepackage{amsmath}
\usepackage{amssymb}
\usepackage{amsfonts}
\usepackage{bm}
\usepackage{color}

\def\be{\begin{equation}}
\def\ee{\end{equation}}
\def\bea{\begin{eqnarray}}
\def\eea{\end{eqnarray}}

\begin{document}

\title{Implosion-explosion in supernovae}
\author{Pierre-Henri Chavanis}
\affiliation{Laboratoire de Physique Th\'{e}orique (UMR 5152 du CNRS),
Universit\'{e} Paul Sabatier, 118 route de Narbonne, 31062 Toulouse Cedex 4,
France}
\author{Bruno Denet}
\affiliation{Universit\'e Aix-Marseille, IRPHE, UMR 7342 CNRS et Centrale
Marseille,
Technopole de Ch\^{a}teau-Gombert, 49 rue Joliot-Curie, 13384 Marseille Cedex
13, France}
\author{Martine Le Berre}
\affiliation{ISMO-CNRS, Universit\'e Paris-Saclay,  91405 Orsay Cedex, France}
\author{Yves Pomeau}
\affiliation{Ladhyx, Ecole Polytechnique, 91128,  Palaiseau, France}

\begin{abstract} 

Supernovae explosions of massive stars are nowadays believed
to result from a two-step process,  with an initial gravitational core
collapse followed by an expansion of matter after a bouncing on the core.  This
scenario meets several difficulties. We show that it is not the only possible
one:  a simple model  based on fluid mechanics and stability properties of the
equilibrium state shows that  one can have also a \textit{simultaneous}
inward/outward motion in the early stage of the instability of the supernova. 
This shows up in the slow sweeping across a saddle-center bifurcation found 
when considering  equilibrium states associated to the constraint of energy 
conservation. We first discuss the weakly nonlinear regime in
terms of a Painlev\'e I equation. We then show that the strongly nonlinear
regime displays a self-similar behavior of the core collapse. Finally, the
expansion of the remnants is revisited as an isentropic process leading
to shocks formation.

\end{abstract}

\pacs{05.00.00, 47.00.00, 97.00.00}

\maketitle

\section{Introduction}
Although supernovae make one of the most spectacular phenomena displayed to us
in the Universe, their understanding remains a challenge. As stated recently in
a review on the subject \cite{Burrows} it is still ``in an unsatisfactory state
of affairs". A major unsolved problem concerns the death of young massive stars,
supposed to occur in two steps, first a collapse then emission of matter and
radiation by an explosion.
 The problem with this picture, relying in part on
 extensive hydrodynamical simulations, is the difficulty to explain the reversal
of the motion, from inward (collapse or {\it{im}}plosion) to outward (the
observed {\it{ex}}plosion of supernovae)  which requires very large outward
forces to turn the tide.
 According to most works on core collapse of supernovae, this reversal is due to
a stiffening of the equation of state at the center,
 which stops the collapse and leads to a bounce. Without bounce, namely if the
shock remains at a more or less fixed location, matter keeps flowing inward and
there is no explosion at all. In numerical studies, an outward propagating shock
can be created, but typically this shock stalls at some radius,  and it is hard
to find a mechanism making it move again outward (see \cite{Burrows,Clavin} and
references herein). Neutrino heating has been invoked but numerical
simulations have shown that this is not generally sufficient to produce an
explosion. More recently, 3D hydrodynamic instabilities have been discussed,
that are still controversial. In summary, the revival of the stalled accretion
shock remains an unexplained process since the early $1980$'s. 

 In our theory, we focus on the pure fluid mechanical part of the physics of
supernovae without considering the immensely complex network of possible nuclear
reactions in the core. In particular, we do not add any mechanism of heating of
compressed matter by increasing the rate of nuclear reactions in the evolving
system, which should take into account the sensitivity of nuclear reactions to
temperature and density, a direct consequence of the fact they are linked to
quantum tunneling. Using the tools of bifurcation theory
  we show that an outgoing velocity field in the early stage of the loss of
equilibrium of the star is compatible with a simple hydrodynamic model. 
In particular, it illustrates that a change from a canonical to microcanonical
description does change the outcome of the evolution: the canonical model
collapses without any outgoing flow as treated in our previous paper
\cite{epje}, although the microcanonical model  studied here shows  this outward
flow! 
In both cases,
 supernova is described as a dynamical catastrophe due to a slow crossing of
an instability threshold by the control parameter,  the
temperature in \cite{epje} and the total energy  of the star here. These
instabilities are manifested by a  loss of balance between the inward pull of
self-gravity and the outward pull of the pressure. The loss of balance is
a \textit{global}  phenomenon that depends on the eigenvalues of
a stability equation  depending itself on the distribution of matter and energy
in the star. Therefore, if the gravity is not dominant everywhere with respect
to the pressure, one expects different orientations of the radial velocity as a
function of the radius in the early post-bifurcation stage, as found here.

Using inviscid compressible fluid equations with gravitation and a particular
equation of state, we showed  in \cite{epje} that the large difference of time
scales between the slow evolution of a star and its quick explosion is explained
by the slow crossing of a saddle-center bifurcation of its steady state. The
dynamics close to the bifurcation is described by a ``universal" Painlev\'e
equation. Later, higher order terms come into play and require a full numerical
study.
In reference  \cite{epje}, the uniform temperature $T$ of the star was the
slowly varying control parameter, a situation which amounts to considering a
star in contact with a thermostat whose temperature decreases very slowly with
time. We refer to this model as the canonical Euler-Poisson model (CEP).
In this case, we did observe a collapse of the whole mass of
the star towards the center, as in published numerical studies of this
phenomenon.
In the present paper, using the same approach as in \cite{epje},
we investigate the same hydrodynamical model, also with spherical symmetry, but
in the microcanonical description (MEP model), namely by adding a condition of
conservation of total energy. In other words, the control parameter is now the
total energy $E$ of the star, supposed to slowly decrease with time.

This change from given $T$ to given $E$
 was motivated by
 previous studies concerning phase transitions in self-gravitating N-body
systems (see the review in \cite{can-microcan}) which may have applications in
astrophysics where galaxies, globular clusters, 
dust gas, fermions gas (like electrons in white dwarfs or neutrons in neutron
stars) are examples of self-gravitating systems. 
 Using thermodynamics and statistical mechanics tools, it was
found
 that very different dynamics characterize canonical ensembles (with fixed
temperature) and microcanonical ensembles (with fixed energy) especially in the
vicinity of phase transitions. In certain situations a collapsed core is formed
in the canonical case, whereas a core surrounded by a halo is formed in the
microcanonical case.
 Therefore, a question naturally sets up: what should be obtained with the fluid
model proposed in \cite{epje} when passing from the canonical description which
leads to a total collapse of the star with a growing singularity at its core, to
the microcanonical one? We show below that we obtain simultaneously a collapse
of the central part of the star, and emission of matter in the outer part.
  This simultaneous inward/outward motion occurs from the early stage of the
bifurcation until the core collapse in the MEP model, but not in the CEP one. 
Indeed, if a constant temperature is assumed, there is an
overall inward motion inside the whole star, from the very beginning of the
loss of equilibrium. If one imposes with the same equation of state the
constraint of energy conservation instead of the one of given temperature,  the
all inward-going velocity field is turned into an inward velocity field near
the center of the star and outward in the rest of the star. This simple model
opens the way to a new understanding of the explosion of stars, based on fluid
mechanics, catastrophe theory, and bifurcation properties of their equilibrium
state.  It also provides
a nice illustration of the property of inequivalence between canonical and
microcanonical ensembles for systems with long-range interactions.

This Letter is organized as follows. We start by describing the
first stage of the dynamics, where nonlinearities with
respect to the amplitude of the instabilities are weak. We derive  the normal
form for the evolution of the unstable amplitude, which takes the generic form
of a Painlev\'{e} I equation, and which agrees with the numerical solution of
the full MEP system. Then, we show that beyond the Painlev\'{e} regime
the numerical solution displays a self-similar behavior of the core collapse,
with scaling laws illustrating, as expected, the dominance of the gravity forces
over the pressure in the core. In this strongly nonlinear phase of the dynamics
the inward  and outward  velocities increase with time, the acceleration of the
inward motion in the core being larger than  the one of the explosion of the
outer shell. Finally, after a short self-similar post-collapse
solution, we propose to interpret the expansion of the remnants as an isentropic
process
described by a Burgers-type equation leading to shocks formation.

\section{Saddle-center in the microcanonical description of a self-gravitating
fluid}
\label{sec:equations}

 The microcanonical model differs from the canonical one presented in
\cite{epje} by an equation expressing the condition of fixed energy which has to
be added. This simple change has noticeable consequences, it modifies the
properties of the equilibrium states, especially the radial profile of the
neutral mode.
Using dimensionless variables such that 
 $G=\rho_*=k_B = M\,=1$ ($G$ and $k_B$ being Newton's and Boltzmann's constants
and $M$ the total mass of the star),
the Euler-Poisson system is
\begin{eqnarray}
\label{e1}
\frac{\partial\rho}{\partial t}+\nabla\cdot (\rho {\bf u})=0,
\end{eqnarray}
\begin{eqnarray}
\label{e2}
\frac{\partial {\bf u}}{\partial t}+({\bf u}\cdot \nabla){\bf
u}=-\frac{1}{\rho}\nabla p-\nabla\Phi,
\end{eqnarray}
\begin{eqnarray}
\label{e3}
\Delta\Phi=4\pi \rho,
\end{eqnarray}
where
${\bf u}({\bf r},t)$ is the radial fluid velocity and $\rho({\bf r},t)$ the
mass density.
We consider a barotropic equation of state of the   
form
$p({\bf r},t)=T(t)g(\rho({\bf r},t))$,
namely with uniform temperature as in \cite{epje}. However, differently from
\cite{epje},
we assume that the temperature $T(t)$ evolves so as to conserve the total
energy, with a simple energetic constraint of the form
\begin{eqnarray}
\label{e5}
E=\frac{1}{2}\int \rho {\bf u}^2\, d{\bf r}+\frac{3}{2}
T(t)+\frac{1}{2}\int\rho\Phi\, d{\bf r},
\label{ae6}
\end{eqnarray}
which determines the uniform temperature $T(t)$ for a given $E$.

 We take the same equation of state as in \cite{epje},
\begin{eqnarray}
\label{i1}
p(\rho, T)=T\left (\sqrt{1+\rho}-1\right )^2,
\end{eqnarray}
which reduces to the isothermal equation $p=T\rho$ 
in the high density region near the core, and to the polytropic equation of
state $p=T \rho^2/4$ with index $2$ at low density (edge of the star). This
equation allows to confine the system in a finite region, while having an
isothermal core. Solution with an isothermal core was chosen because previous
studies on isothermal spheres enclosed in a box have led to a saddle-center
bifurcation \cite{emden,aaiso} as the energy or the temperature varies.

Our first step is to investigate whether such a transition exists within our
model. To do that we compute the equilibrium states, which are conveniently
obtained  from the enthalpy $h$ defined by the relation $dh=dp/\rho$,
 or $h(\rho)=\int_0^{\rho} \frac{p'(\rho')}{\rho'} \, d\rho$.  Adding the
relation $h(\rho=0)=0$ (which defines the radius $r_{0}$ of the star), one
obtains
 $h(\rho, T)=2  \ln \left ( 1+\sqrt{1+\rho}\right )-2 \ln (2)$.

\begin{figure}
\centerline{
(a)\includegraphics[height=1.0in]{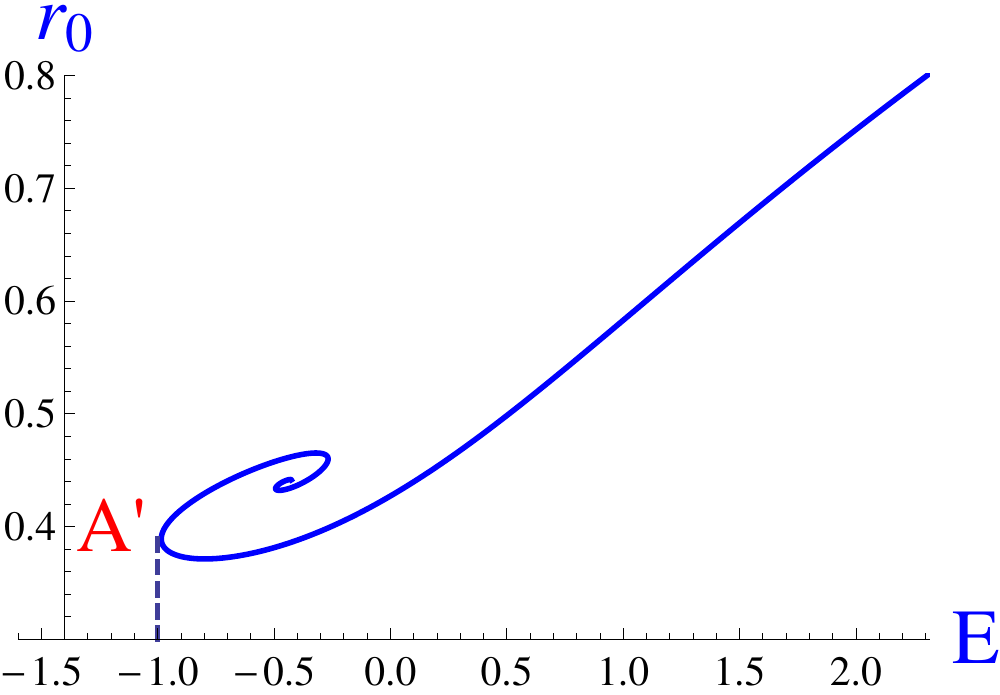}
(b)\includegraphics[height=1.0in]{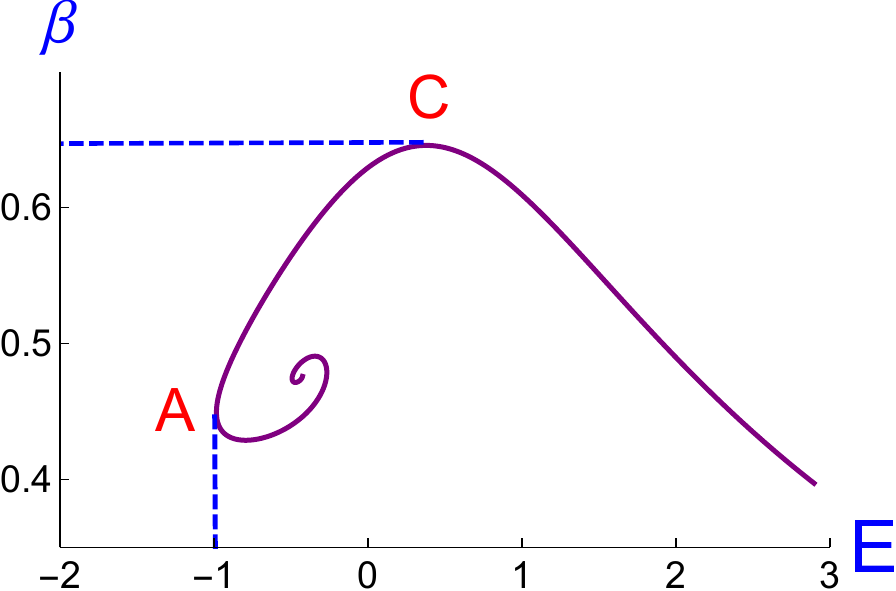}}
\caption{
  Equilibrium state radius versus energy $r_0(E)$ in (a);  caloric curve
$\beta(E)$ in (b) with $\beta=1/T$. The spiral in (b) displays the two critical
points $A$ and $C$ which characterize the first bifurcation 
occurring respectively at $E_c\simeq -0.984$  for the MEP model, and at
$\beta_c^{\rm cano}\simeq 0.647$  for the CEP model.
 }
\label{fig:spiA}
\end{figure}

\begin{figure}
\centerline{
(a)\includegraphics[height=1.5in]{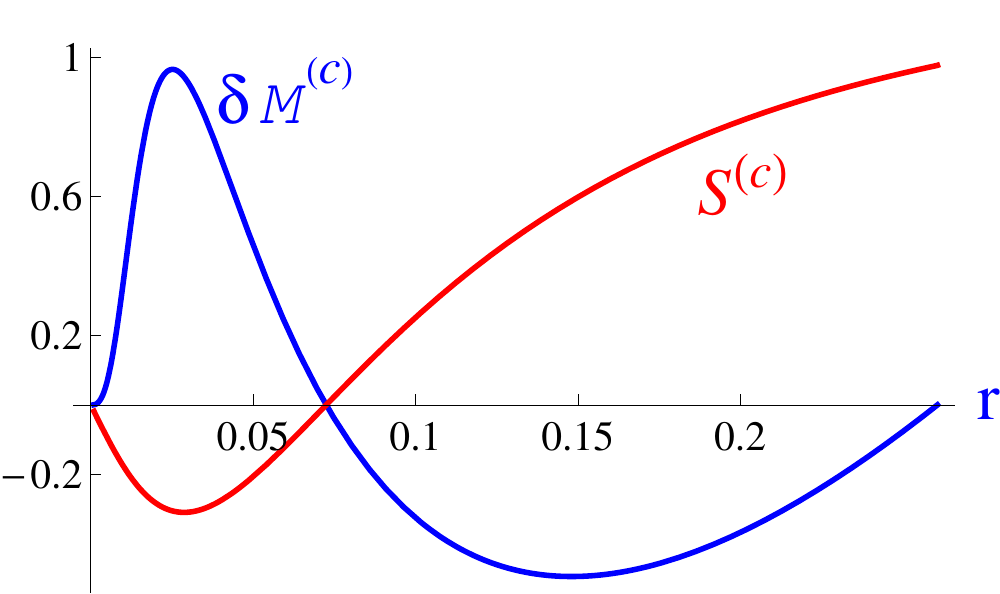}
}
\centerline{
(b)\includegraphics[height=1.5in]{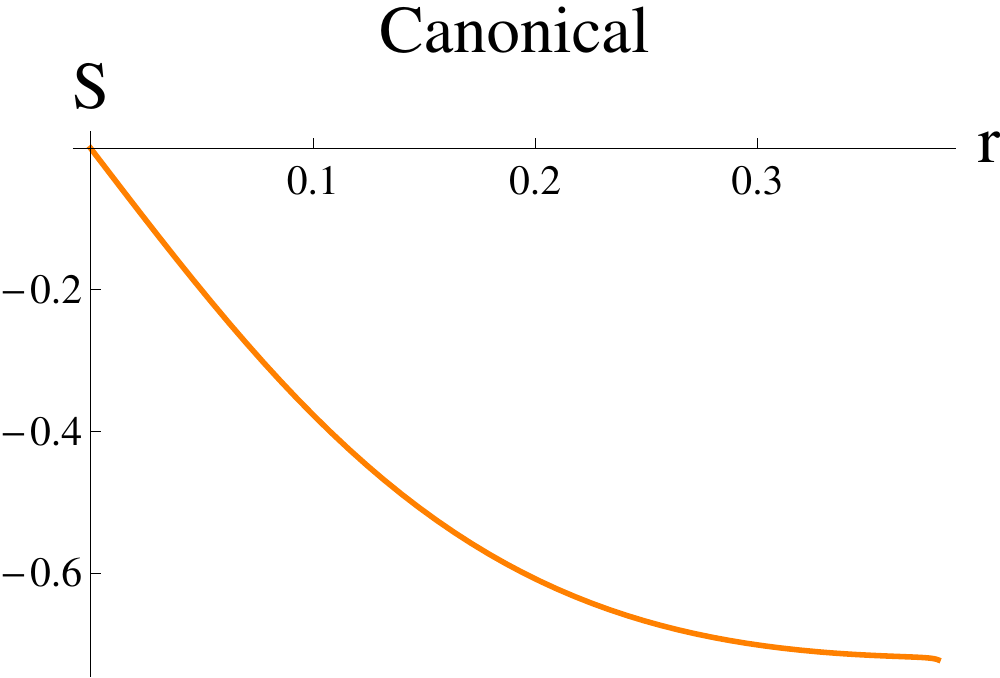}}
\caption{
(a) MEP model: first order radial profiles for the neutral mode of the velocity
(or displacement) $S(r)$  and of the mass deviation $\delta M^{(c)}(r)$ 
 close to the saddle-center.
(b)  CEP model: neutral mode profile of the velocity.
}
\label{fig:neutral}
\end{figure}

Using the  set of dimensionless variables and fields as
defined  in \cite{epje} in order to have a single free parameter, $\hat{h}_0=
h(r=0)/T$,  varying the value of $\hat{h}_0$ and returning to the above
dimensionless variables, we can draw the spiralling curves depicting the steady
states, such as the radius of the star $r_0(E)$ or the inverse temperature
$\beta(E)$ shown in Fig. \ref{fig:spiA}. In curve (a), following the spiral from
large values of $E$ until the lowest value $E_c \simeq -0.984$, one can show
that the solution is linearly stable before the turning point $A'$ and becomes
unstable after. The point $A'$ is a saddle-center for the microcanonical
description because here a stable center merges with an unstable saddle.
 Curve  (b) displays the caloric curve $\beta(E)$ with its 
 two critical points $C$ and $A$ characterizing the first bifurcation for the
CEP and MEP models respectively (see the Figure caption). In between $A$ and
$C$ (i.e., in the region where the specific heat $C=dE/dT$ is negative)
the system is unstable for the CEP model, whereas it remains stable for the MEP
model.

\section{Painlev\'e regime} Close to $A$ the weakly nonlinear analysis detailed
in \cite{art-long} is performed by expanding the MEP solution in powers of a
small parameter $\epsilon$ associated to the slow decrease of the energy of the
form $E(t)=E_c-\gamma' t$, written as
\begin{equation}
E=E_c-\epsilon^2 E^{(2)},
\label{n1}
\end{equation}
that amounts to defining $\epsilon^2 E^{(2)}=\gamma' t$ and taking  $\epsilon$
small.

At first order, setting $\delta M(r,t)=  \epsilon
{A}^{(1)}(t)\delta M^{(c)}(r)$ and similar expressions for other small
variations, one finds an ordinary integro-differential equation for the neutral
mode profiles.
 The radial profiles $S(r)$ of the velocity field and mass deviation $\delta
M^{(c)}(r)=\int_0^{r}\delta \rho^{(c)}(r') 4\pi {r'}^2\, dr'$ shown in
Fig. \ref{fig:neutral}-(a)  illustrate well the joint inward/outward motion of
matter in this small amplitude regime. The velocity is negative (inward
motion) in the core and positive (outward motion) in the halo. The important
point is that the double
direction occurs simultaneously during the early stage of the bifurcation, as
observed numerically.  For comparison, we present in Fig. \ref{fig:neutral}-(b) 
the neutral mode velocity profile of the CEP model, which clearly shows an
inward motion everywhere in the star, as confirmed by the numerics \cite{epje}.
This is the key point of the present study, illustrating the very different
behavior of the solution from the very beginning of the approach to the phase
transition, whether we consider a model including the energetic constraint
(\ref{ae6}) or not.

\begin{figure}
\centerline{
 \includegraphics[height=1.75in]{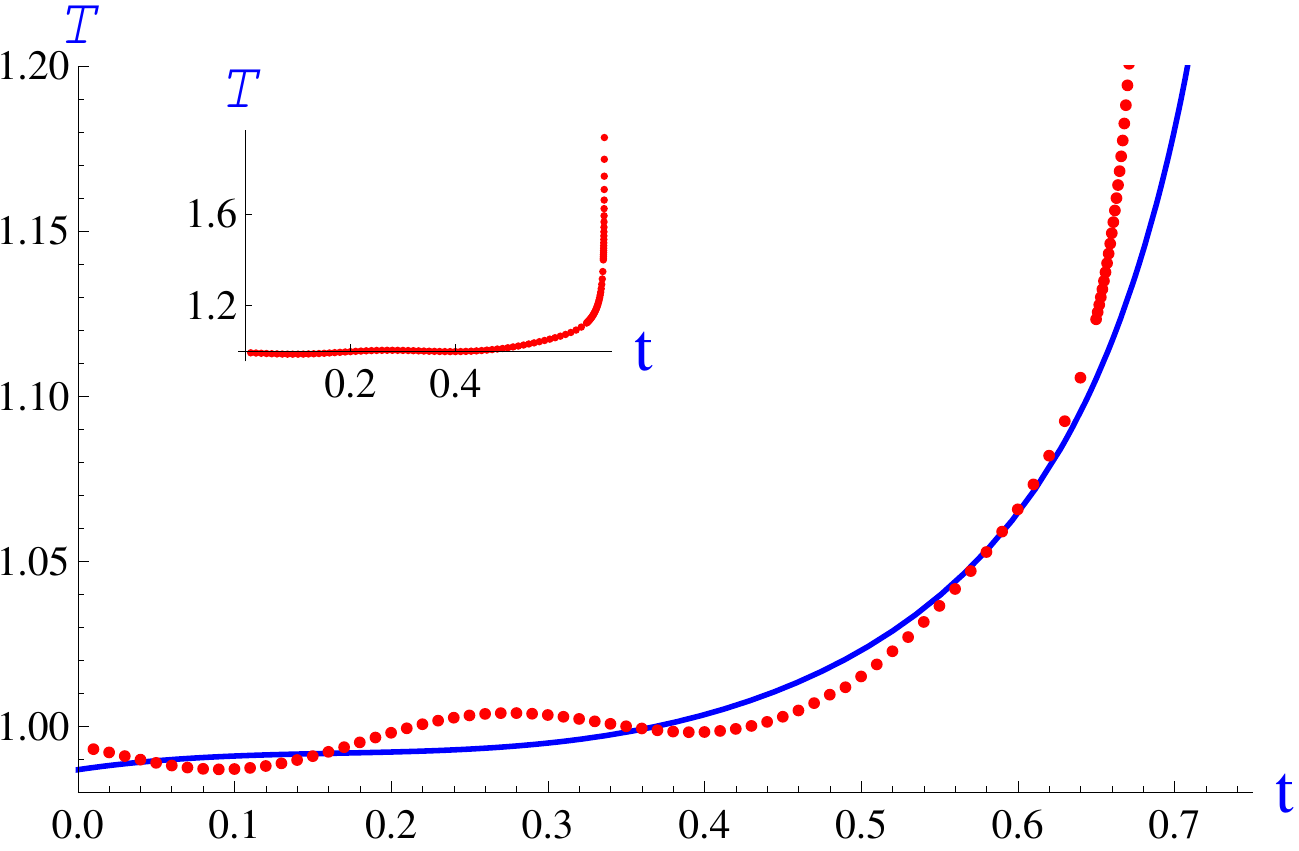}
}
\caption{
Temperature evolution: Full numerical solution (dots) of MEP equations with the
energetic constraint $E=E_c-\gamma' t$ and $\gamma'=0.1$ compared to the
solution $A(t)$ of Painlev\'{e}  equation (\ref{eq:Painleve}).
  The insert displays the numerical solution on a longer time interval, from the
equilibrium stage until the collapse/explosion of the star.
}
\label{Fig:Painl}
\end{figure}

At order two, the
solvability condition amounts also to solving an integro-differential equation,
which leads ultimately to a normal form of Painlev\'{e} I type,
\begin{equation}
\ddot{{A}}(t)=   \gamma t + K {A}^2
\mathrm{,}
\label{eq:Painleve}
\end{equation}
for the time evolution of the deviations close to the critical point. We find
$\gamma= 46.63.. \gamma'$ and $ K = 1055.98..$,
that gives the temperature evolution drawn in Fig. \ref{Fig:Painl}, solid line, 
in good agreement with the numerical results (red dots) of the full MEP
equations for this early stage of the implosion-explosion process. For
comparison, we note that for the CEP model, the normal form (also of
Painlev\'{e} I type)  was found with coefficients $\gamma\simeq 120.2$ and
$K\simeq 12.3$, leading to a slower growth of the amplitude. This relies on the
fact that the critical density at $r=0$ is much lower (by a factor $100$) for
the CEP model.

\section{Post-Painlev\'{e} dynamics (pre-collapse)}
\label{sec:post}

\begin{figure}
\centerline{
(a)  \includegraphics[height=2in]{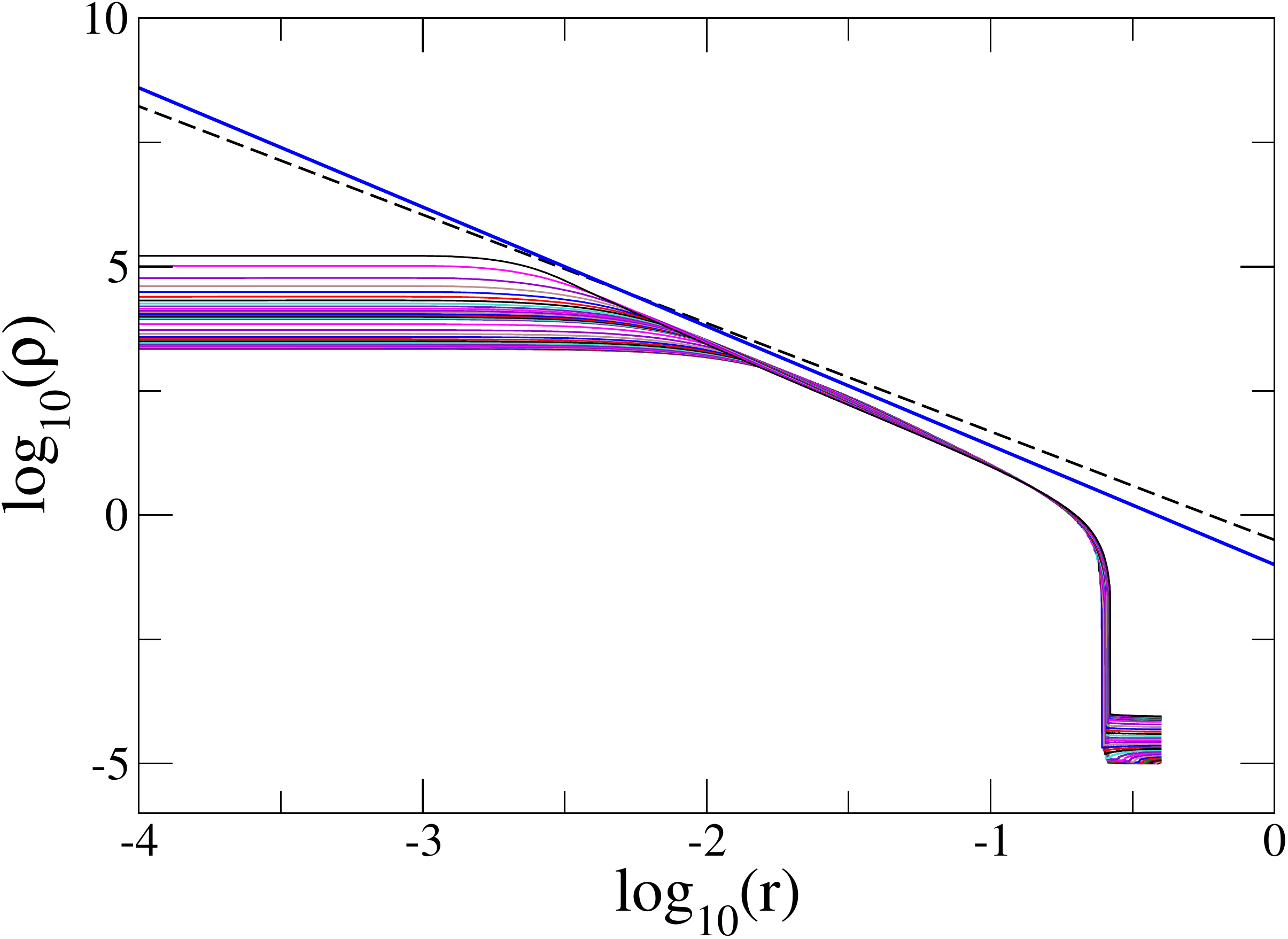}}
\centerline{  
(b) \includegraphics[height=2in]{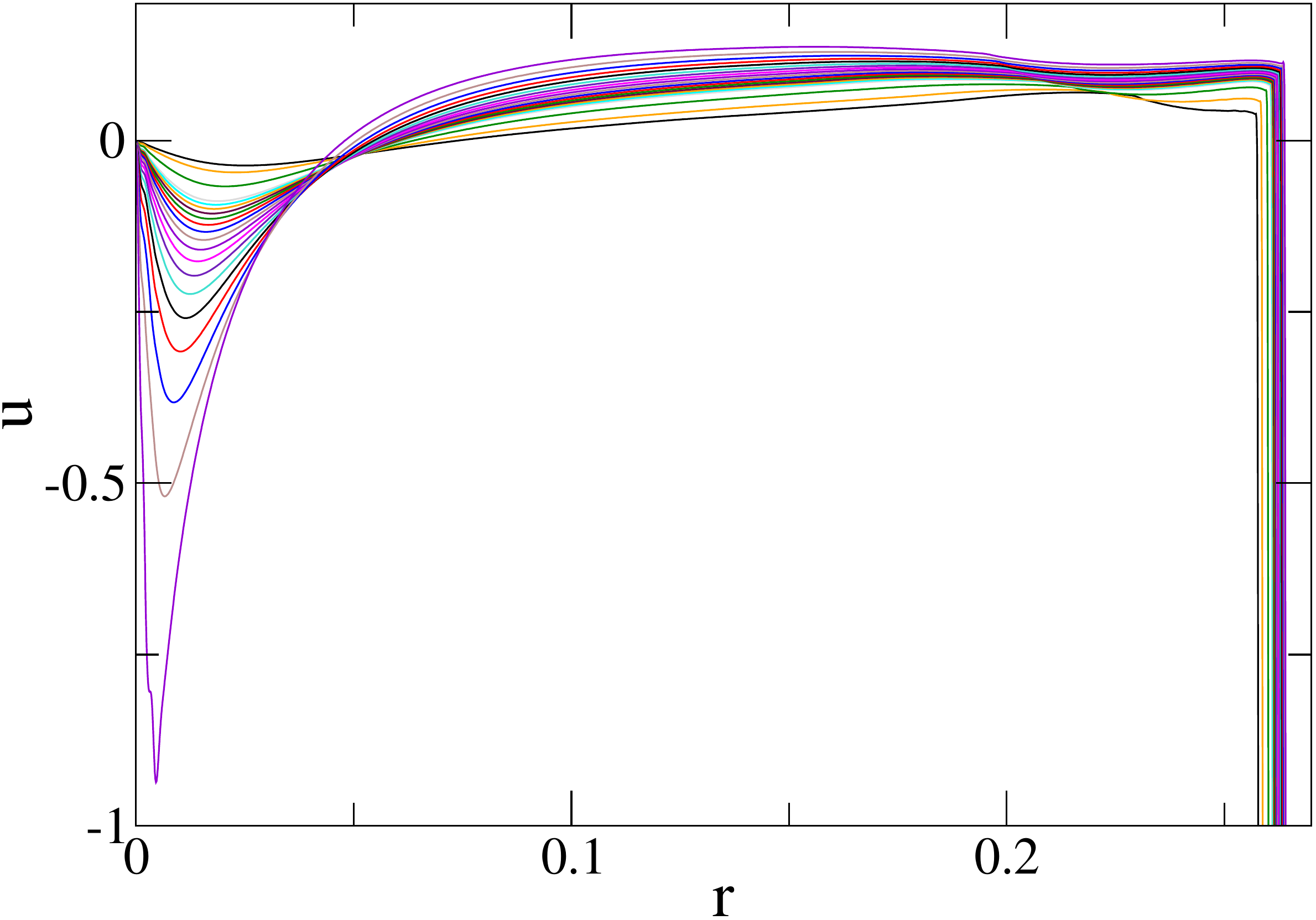}
}
\caption{
 Numerical solution of MEP equations at successive time values in the
post-Painlev\'{e} stage. Left: density  versus radius in $\log_{10}$ scale.
Right: velocity  versus radius. The modulus of the extrema increase 
with time in both curves.
}
\label{Fig:post-Painl}
\end{figure}

After the weakly nonlinear Painlev\'{e} regime, 
the full MEP system of equations displays a solution which ultimately diverges,
as illustrated in the insert of Fig. \ref{Fig:Painl}. This divergence of the
temperature is associated to core-collapse, as illustrated in
Fig. \ref{Fig:post-Painl} where the density and velocity are drawn versus $r$
for successive time values (the maxima of the density and absolute velocity
profiles increase with time). In (a) the logarithmic scale shows that the
solution in the core becomes self-similar. The singularity is of second kind in
the sense of Zel'dovich, as already found in \cite{epje} for the CEP model
leading to the collapse of the whole star. For both cases (MEP and CEP models)
the core collapse is characterized by the fact that gravity dominates over
pressure forces, but the exponents found here are different from the previous
ones. Recall that for the gravity-dominating case, using the notations of
\cite{epje} the self-similar density is of the form
 \begin{equation}
\rho(r,t) = (-t)^{-2} R(r(-t)^{-2/\alpha})
\mathrm{,}
\label{eq:rhoa}
\end{equation}
and the velocity
 \begin{equation}
 u(r,t) = (-t)^{-1+\frac{2}{\alpha}} U(r(-t)^{-2/\alpha})
 \mathrm{,}
\label{eq:ua}
\end{equation}
where
 $\xi=r(-t)^{-2/\alpha}$. That gives
$R(\xi)\sim \xi^{-\alpha}$, and $ U(\xi)\sim \xi^{-(\alpha/2-1)}$ for
$\xi\rightarrow +\infty$ in order to have a steady profile at large distances.
The  exponent $\alpha$ is related to the behavior of the self-similar solution
as $\xi$ tends to zero \cite{epje}. More precisely, expanding $R$ as $R = R_0 +
R_k \xi^k+ ..$ (and $U$ as  $U= U_1 \xi  + U_k \xi^{k+1}  + ...$), one has 
$\alpha(k)=\frac{6k}{2k+3}$. In the present case we find numerically that the
asymptotic behavior of the density displays an exponent $\alpha$ larger than
two, that is equivalent to the gravity dominance (over pressure) property. 
More precisely, we obtain a best fit  (solid line) with the value 
 \begin{equation}
\label{eq:alpha}
\alpha=48/19
\end{equation}
 which corresponds to $k=8$, namely to the on-axis behavior $R = R_0 + R_8
\xi^8+... $ 
For comparison, we plot in dashed black line the slope corresponding to the
value $\alpha=24/11$, or $k=4$, found for the CEP model \cite{epje}. 

In this strongly nonlinear regime standing from the Painlev\'{e} (weakly
nonlinear) stage ending at $t_s$ up to the divergence of the solution at $t_*$,
the absolute value of the
inward \textit{and} outward velocity increase everywhere in the star, as
illustrated in Fig. \ref{Fig:post-Painl}-(b).  The velocity curves clearly
show that the inward motion accelerates more rapidly (in the core) than the
outward motion taking place in 
the outer shell. This remark allows us to deduce the
temperature behavior in the post-Painlev\'e regime.  For $t_{s} < t < t_{*}$,
using Eq. (\ref{ae6}) and neglecting the contribution of the halo in the energy,
we get \cite{art-long} $T(t)\propto (t_*-t)^{(15-2k)/3k}$, hence for $k=8$,
\begin{equation}
T(t)\propto (t_*-t)^{-1/24},
\end{equation}
which diverges at the collapse time.
During the self-similar growth of density and inward velocity in the core, what
happens in the halo?  From Fig. \ref{Fig:post-Painl}-(a) we observe by zooming
on the halo region that the density decreases with time. Moreover,
Fig. \ref{Fig:post-Painl}-(b) clearly shows that the velocity which diverges in
the core in the pre-collapse regime,  barely increases and remains finite  in
the halo which expands radially by about $20$\%, a small evolution compared to
the strong shrinking of the core. 
Investigating possible self-similar solutions of the form
(\ref{eq:rhoa}) and (\ref{eq:ua}) for the halo,  it is shown in \cite{art-long}
that no self-similar solution, if it exists, agrees with  the
numerical curves,
either assuming that gravity is negligible with respect to the pressure, or the
inverse, or else in the mixed case. 
We conclude that no self-similar solution is able to describe 
the dynamics of the halo before the collapse,  the expansion of the halo being
still in a preliminary stage.

\section{Post-collapse}
Just after the collapse time when the mass at the center 
of the star is zero, self-similar solutions exist  in the core and in the halo,
as detailed in \cite{art-long}, and summarized below. They are obtained under
the hypothesis that gravity overcomes pressure forces in the core, and the
opposite in the halo. 
In the core, the  post-collapse situation is then qualitatively the
same as in the CEP model, and  looks (mathematically) like the one of the
dynamics of the Bose-Einstein
condensation where the mass of the condensate begins to grow from zero
{\it{after}} the time of the singularity
\cite{BoseE,bosesopik}.  For the MEP model, the self-similar
solution  is the one derived in \cite{epje} but with the value of the
exponent $\alpha$ found here, Eq. (\ref{eq:alpha}). 
We recall that the main change with respect to the pre-collapse study amounts to
adding to the equations of density and momentum
conservation, an equation for the mass at the center $M_c(t)$ (with $M_c(0) =
0$), 
\begin{equation}
\frac{d{M}_{c}}{dt} = \left[-4 \pi r^2 \rho(r) u(r)\right]_{r \to 0}
\mathrm{.}
\label{eq:Mc}
\end{equation}
We get $M_{c}(t) \propto t^{b}$,  with $b=6/\alpha-2$ a 
positive exponent \cite{epje}. Because the scaling laws are the same before and
after the
singularity, Eq. (\ref{eq:alpha}) yields  $b=3/8$. If we replace the
singular core by a relativistic compact object such as a neutron star, we get an
estimate of the temperature evolution by the relation
 $k_B T\sim M_c\, c^2$ leading to the scaling
\begin{equation}
T(t) \propto  t^{{3}/{8}}.
\label{tev}
\end{equation}

In the description of the halo expansion just after the explosion, we assume
that the energy  released  during the collapse of the core heats the halo and
provides its expansion. Indeed, as the
gravitational energy $W$ of the core decreases and becomes very negative, the
temperature $T$ of the halo given in (\ref{tev})  and its macroscopic kinetic
energy $E_{\rm kin}$
increase and become very large
($T\sim E_{\rm kin}\sim -W$) as a result of energy conservation.
Therefore, the pressure inside the halo can be high enough to accelerate its
expansion. We assume that  the pressure forces in the halo are stronger than the
gravity forces, a condition  checked {\it in fine}, and also that the velocity
increases linearly with the radius,
$u(r,t)=H(t)r$.  Within this frame, and for a polytropic equation of state
$P=K(t)\rho^{\gamma}$, where $K(t)$ increases with time, we show in Appendix C
of  \cite{art-long}  that the radius $R(t)$ of the halo is solution of 
\begin{eqnarray}
\label{eq:phc2}
\ddot R R^{3\gamma-2}  = K(t).
\end{eqnarray} 
The  expansion rate $\dot R(t)$
is time dependent, a
result which differs from the common
description of the
remnant motion just after the explosion (supposed to expand with a constant
velocity due to the conservation of kinetic energy). 
For $\gamma=1$ and $T(t) \propto  t^{{3}/{8}}$, the radius increases as
$R(t)\propto t^{19/16}$, a solution that is expected to merge ultimately with
the non self-similar Burgers solution suggested just below, which displays
shocks.

\section{Free expansion stage (shocks)} The last stage of the evolution of the
supernova explosion
is of interest, although it  is often considered as  rather uneventful. In this
so-called ``free expansion
stage''  the ejecta are generally believed to
cool adiabatically because of their free expansion.  
 We consider remnants as making a dilute gas, much denser than the interstellar
medium. This free expansion
stage  lasts until the density of the remnants becomes of the order of
magnitude of the one of the interstellar gas, so the expanding gas
is dense enough to be interacting with itself (the attraction by the
core of the exploded star is  also considered), but not with the
interstellar medium, and  it makes a continuous fluid, not a Knudsen
gas. Our argument relies on the fact that the mean-free path  in
interstellar matter may be as large as the size of a galaxy, that makes
 such event unrealistic. In that frame,
the remnant is an entity which does not exchange
energy and mass with the interstellar medium. Therefore,  we assume that the
expansion of a gas {\textit{bubble in vacuo}}  is an isentropic process with two
constraints, the conservation of mass and energy.  We start from the equations 
for an inviscid compressible  ideal fluid with spherical symmetry,  see section
6 of \cite{ll},  which include the isentropic condition
\begin{equation}
 \frac{\partial (r ^2 s \rho)}{\partial t} + \frac{\partial}{\partial r}(u r^2 s
\rho) = 0  
\textrm{,}
\label{eq:entropy}
\end{equation}
where $s$ is the entropy per unit mass.  The pressure $P$ is now a 
function of $(\rho,s)$, $P=K(s) \rho^{\gamma}$,  with $\gamma$ larger than
unity. Assuming that entropy is initially uniform, we try to find a possible
self-similar solution  with spherical symmetry  of the form  $F(r, t) = r^a f(r
t^b)$, where the exponent 
 $a$ depends
on the field $F$ under consideration (that is either $ \rho$, $u$ or $s$)
although $b$ is the same for all fields.    

 First, neglecting the gravity with respect to the
pressure forces, we find that  the two conservation laws impose that   $a=-3$,
$b=-1$, and $\gamma=1$. The latter  condition  is incompatible with the
definition  of $\gamma=c_{p}/c_{v}$. We conclude that a self-similar expanding
solution for the halo is not physically meaningful, contrary to the
free fall of dense molecular gas (where the opposite was assumed). 

Secondly, we assume that  gravity {\textit{and}} pressure  forces can be
neglected, that reduces the momentum equation to
\begin{equation}
\frac{\partial u}{\partial t} + u \frac{\partial u }{\partial r} = 0
\textrm{,}
\label{eq:momentmod}
\end{equation}
an equation well-known since Poisson to have the implicit solution
\begin{equation}
u(r, t) = u_0 (r - u t)
\textrm{,}
\label{eq:momentmodsol}
\end{equation}
where $u_0(r)$ is the initial radial velocity. This solution conserves
the order of magnitude of $u$ in the course of time, which is consistent with
the conservation of energy.  
This Burgers-type equation is a prototype for creating shocks. The solution for
the density is
\begin{equation}
  r^2 \rho(r, t) = \frac{\partial r_0 }{\partial r}   r_0^2 \rho_0(r_0(r, t)) 
\textrm{,}
\label{eq:momentmodsoldens}
\end{equation}
where the index $0$ refers to the initial condition, as above. We check
\cite{art-long} that the  term $u \partial u /\partial r$ in Eq.
(\ref{eq:momentmod}), the dynamical pressure, is dominant over the term of
thermodynamical pressure, $P_{,r}/\rho$, and over the gravitational term $- 
(4\pi G/r^2) \int_0^r  \mathrm{d}r' 
r'^2 \rho (r')$, during the expansion of a dilute gas before it enters into the
Knudsen  regime.
Shocks  forms  if the radial velocity is larger for a
given radius than for a larger one, because the larger velocities
overcome the slower ones. Therefore, shocks are formed naturally
inside the remnant,  depending on the initial distribution of the fluid velocity
{\it inside} the remnant, and they propagate inside this matter, the role of
the interstellar medium being ignored.

This study shows
that  in order to create structures in an expanding gas volume, as observed in
the remnants, there is no need to have interaction with an outside interstellar
gas. This early stage of the expansion is, by far,
the one that is the best known experimentally because it is a stage where the
remnants are still far more luminous than the rest of the Galaxy. Based on the
existence of such internal shock waves, we suggest an explanation for the very
sharp luminous rings observed in the remnants of SN1987A. 

Finally, we have to note  that we have neglected a set of perhaps crucial
phenomena, namely plasma effects due
to the finite electric conductivity of the expanding gas (this yields Laplace
forces which could supersede inertia and gravity in the expanding gas). We have
 also assumed spherical symmetry, not displayed by the observed
remnants, except for their large-scale structure, but asphericity is
not so crucial from the point of view of the present analysis.

\section{Discussion}
\label{sec:discussion}
Presently, theories of supernova explosion focus on a physical phenomenon, the
emission of neutrinos, or on 3D effects which we do not consider at all in our
work.
In our theory, we focus on an entirely different part of the complex physics of
supernovae; 
namely the fluid mechanical part.
We show that implosion \textit{and} explosion taking place at the death of a
massive star may occur simultaneously. This yields an alternative explanation to
the yet unsolved problem of supernovae description where the two steps process
is an unsatisfactory explanation. Using a simple model,
we point out that the huge difference of time scales between the long life of a
star and its abrupt death can be understood in the light of catastrophe theory,
by a slow sweeping of a saddle-center bifurcation. Starting from the stable
equilibrium state and approaching the saddle-center, we show that the weakly
nonlinear analysis leads to a universal Painlev\'{e} equation for the amplitude
of a bi-modal motion (with two opposite directions).  Our study  illustrates
once more (see \cite{can-microcan}) that a change from canonical to
microcanonical description, not looking very important, does deeply change the
outcome of the transition from stable to unstable state. Here the MEP model
shows an explosive outer shell together with a core collapse although the CEP 
model studied in \cite{epje} shows a collapse of the star
without any outgoing flow. This is a manifestation of
ensembles inequivalence for systems with long-range interactions. Note that the
CEP model could describe most of
hypernovae since supermassive stars often collapse with emission of very intense
gamma ray bursts, but without any explosion of the outer shell.

It is important to point out that  the Painlev\'{e} 
analysis yields a definite sign for the velocity field at critical, contrary to
what happens in ``classical" transition from linearly stable to linearly
unstable situation (where the unstable mode may have either positive or negative
amplitude). This fair property of the definite sign of the growing Painlev\'{e}
solution comes from the fact that in the case of a saddle-center bifurcation,
the two stable and unstable  equilibrium states merge at the critical point,
beyond which no equilibrium state exists (neither stable nor unstable) that
makes the difference with the ``classical" case of a generic instability.
Then, we describe the self-similar core-collapse regime where 
the acceleration is larger in the collapsing core than in the exploding
envelope. 
Finally, after a short self-similar post-collapse solution, we propose to
interpret the expansion of the remnants as an isentropic process which conserves
the energy and the mass of the halo. Within this  description, there is no
self-similar solution, contrary to the free fall of dense molecules, but another
type of solution appears, of Burgers-type, which is a prototype for creating
shocks. In our rough description, shocks are formed naturally
inside the remnant, and they propagate inside this matter, the role of the
interstellar medium being ignored.

\end{document}